\def\grs{GRS~$1915$+$105$}
\def\X1550{XTE~J$1550$--$564$}
\def\ergcms{erg~cm$^{-2}$~s$^{-1}$ }
\def\integral{{\it{INTEGRAL}}}
\def\rxte{{\it{RXTE}}}

\documentclass[]{aastex}
\usepackage{emulateapj5}

\usepackage{amssymb}
\usepackage{natbib}
\usepackage{epsfig}
\usepackage{epsf}
\usepackage{color}
\definecolor{red}{rgb}{0.7,0,0}
\definecolor{blue}{rgb}{0,0,0.7}
\def\correc#1{{#1}}

\shorttitle{An INTEGRAL monitoring of GRS 1915+105: Part 2}
\shortauthors{Rodriguez et al.}
\begin{document}
\title{Two Years of INTEGRAL monitoring of GRS 1915+105\\
Part 2: X-Ray Spectro-Temporal Analysis}

\author{J. Rodriguez\altaffilmark{1}, S.E. Shaw\altaffilmark{2}, D.C. Hannikainen\altaffilmark{3},  T. Belloni\altaffilmark{4}, S. Corbel\altaffilmark{1}, M. Cadolle Bel\altaffilmark{5}, J. Chenevez\altaffilmark{6}, L. Prat\altaffilmark{1},  P. Kretschmar\altaffilmark{5}, H.J. Lehto\altaffilmark{7},  I.F. Mirabel\altaffilmark{8},  A. Paizis\altaffilmark{9},  G. Pooley\altaffilmark{10},  M. Tagger\altaffilmark{11}, P. Varni\`ere\altaffilmark{12}, C. Cabanac\altaffilmark{2}, O. Vilhu\altaffilmark{3}}

\altaffiltext{1}{Laboratoire AIM, CEA/DSM - CNRS - Universit\'e Paris Diderot, DAPNIA/SAp, F-91191 Gif-sur-Yvette, France}
\altaffiltext{2}{School of Physics and Astronomy, University of Southampton, SO17 1BJ, UK}
\altaffiltext{3}{Observatory, PO Box 14, FI-00014 University of Helsinki, Finland}
\altaffiltext{4}{INAF-Osservatorio Astronomico di Brera, via Bianchi 46, 23807 Merate, Italy}
\altaffiltext{5}{European Space Astronomy Centre (ESAC)
Apartado/P.O. Box 78, Villanueva de la Ca\~nada, E-28691 Madrid, Spain}
\altaffiltext{6}{Danish National Space Center, Technical University of Denmark, Juliane Maries Vej 30, 2100 Copenhagen, Denmark}
\altaffiltext{7}{Tuorla Observatory and  Department of Physics, University of Turku V\"ais\"al\"antie 20, FI-21500 Piikki\"o, Finland}
\altaffiltext{8}{European Southern Observatory, Chile. On leave from CEA-Saclay, France}
\altaffiltext{9}{IASF Milano-INAF, Via Bassini 15, 20133 Milano, Italy}
\altaffiltext{10}{Astrophysics, Cavendish Laboratory, J.J. Thomson Avenue, Cambridge CB3 0HE, UK}
\altaffiltext{11}{Service d'Astrophysique (UMR AstroParticules et Cosmologie), CEA Saclay 91191 Gif-sur-Yvette, France}
\altaffiltext{12}{LAOG, Universit\'e J. Fourier (UMR5571), Grenoble, France}

\begin{abstract}
This is the second of two papers presenting the results of two years of 
monitoring of \grs. We focus here on 
the results obtained at X-ray energies with \integral\ and \rxte\ and in the radio with
the Ryle Telescope at 15~GHz. 
We present the X-ray spectral and temporal 
analysis of four observations which showed strong radio to X-ray correlations.
During one observation \grs\ was in a steady state, while during the 
 three others it showed cycles of X-ray dips and spikes (followed by 
 radio flares).  We present the time-resolved spectroscopy
of these cycles with the view to understand the X-ray to radio connection.  
Our spectral analysis shows that in all cases 
the hard X-ray component, which we interpret as the Comptonized emission 
from a coronal medium, is suppressed in coincidence with a soft X-ray spike
that ends the cycle. We interpret these results as  evidence that the soft 
X-ray spike is the trigger of the ejection,  and that the ejected medium is the 
coronal material.  In addition, we study the rapid variations,  and report the 
presence of a low frequency QPO with variable frequency during all X-ray 
dips observed by \rxte. The ubiquity of 
the LFQPO during the dips, prior to the ejection, may indicate a physical link
between those phenomena. In the steady state observation, the X-ray  
spectrum is indicative of the hard-intermediate state, with the 
presence of a relatively strong emission, above $40$~mJy at 15 GHz. The X-ray spectra 
are the sum of a Comptonized component and an extra power law extending to energies 
$>$200~keV without any evidence for a cut-off. We observe 
a possible correlation of the radio flux with that of the power law component, which
 may indicate that we see direct emission from the jet at hard X-ray energies.
 We also present the energy dependence of a $\sim4$ Hz Quasi-Periodic Oscillation
of constant frequency present during the \correc{hard-intermediate state} observation. This QPO-``spectrum''
is well modeled by a power law with a cut-off at an energy about $11$~keV. 
We show that the energetic dependences of 
the QPO amplitude and the relative contribution of the Comptonized component to the 
overall flux are clearly different. This may rule out models of global oscillations 
of the Compton corona.
\end{abstract}
\keywords{accretion, accretion disks --- black hole physics --- stars: individual 
(GRS 1915+105, XTE J1550$-$564) --- X-rays: stars}

\section{Introduction}
\indent X-ray binaries and microquasars are known to be sources of copious 
X-ray emission from a few hundred eV up to the MeV range in some cases.
This high energy emission is commonly attributed to a thermal accretion disk 
for the softer X-rays (below approximately $10$~keV), while no real consensus has 
yet been found for the so-called hard X-ray tails that can be detected up to 
the MeV range \citep[e.g.][]{grove98}. This emission is  usually attributed to an inverse 
Compton process of the soft (disk) photons on a distribution of electrons thought to 
represent a corona whose geometry is a matter of debate. The electrons can have a Maxwellian 
distribution of their velocity (thermal Comptonization), or not. They can represent an 
hybrid population  \citep[thermal and non-thermal Comptonization, e.g.][]{pouta96}, or 
 be in free falling inflow with bulk motion \citep[e.g.][]{laurent99}. 
Alternatives to pure Comptonization exist. In states showing the presence 
of an optically thick compact jet, the high energy emission could also 
originate from Synchrotron Self-Compton (SSC) radiations from the jet 
\citep[e.g.][]{markoff05}. The relative strength of the spectral components together
with the behavior of microquasars in the temporal domain at frequencies typically 
$>0.1$ Hz led to the identification of the so-called spectral states 
\citep[e.g.][for recent definitions]{remillard06,homan06}. Following the 
changes of spectral states in these sources using a multiwavelength approach
is the only way to probe the physics of accretion and its connection with the jets.\\
\indent In this respect, GRS~1915+105 is probably one of the best microquasars to perform 
such studies. It is one of the brightest X-ray sources in the sky and 
it is a source of superluminal ejections \citep{m&r94}, with true velocity of 
the jets $\geq 0.9$c. It is also known to have a strong compact jet during its periods 
of ``hard'' and steady X-ray states \citep[e.g.][]{fuchs03}, while multiwavelength
approaches  have shown a clear connection between the presence of soft X-ray dip/spike 
sequences (hereafter 
X-ray cycles) followed by plasma ejections \citep{mirabel98,fender98}. An complete review 
on \grs\ can be found in \citet{fender04a}.
 \grs\ has been extensively observed with the {\it Rossi X-ray Timing Explorer} ({\it
RXTE}) since 1996.  A rich pattern of variability has emerged from
these data with time scales ranging from years down to 15~msec \citep[e.g.][]{morgan}.  
\citet{belloni00},  classified all the observations into
12 separate classes, which could be interpreted as transitions between
three basic spectral states: a hard intermediate one \correc{(State-C)}, and two soft 
intermediate ones \correc{(State-A and State-B)} following the classification of \citet{homan06}. 
\correc{Note that State-A is much dimmer and softer than State-B that is the brightest state of 
all three.}\\
\indent This is the second of two papers reporting the results of two years of 
an extensive monitoring campaign involving several instruments, namely \rxte, the Ryle 
Telescope (RT), and \integral, the latter being the main 
observatory of our campaign. In Rodriguez et al. 2007 (hereafter Paper~1) we presented the 
results of the multiwavelength analysis. In particular, we focused on four \correc{particular} 
observations. \correc{These four observations were choosen because of the availability of 
simultaneous radio and X-ray observations. In all four of them the 
radio to X-ray connections could then be studied in great details, and showed interesting links.}
In three of them we found clear associations between the X-ray behavior of the source 
and the occurrence of radio flares. In the fourth one, the level of radio emission was 
rather steady and indicative of a compact jet. 
In Paper~1, we generalized the fact that (non-major) discrete ejections 
always occur as a response to an X-ray cycle composed of an X-ray dip longer than 100~s 
ending in an X-ray spike. Based on a model-independent analysis, we also suggested that the 
spike was the trigger of the ejection. We also found a possible correlation between 
the duration of the X-ray dip and the amplitude of the radio ejection, which may 
suggest that energy and/or matter accumulated during the dip, are later ejected. \\ 
\indent In this paper we focus on the X-ray spectral and temporal analysis of 
those four particular observations.  We start by  describing 
the  methods of data reduction in Section 2. In Section 3, 
we present the results, which are discussed in the last part of the paper.\\

\section{Observations and Data Reduction}
The log of all observations can be found in Paper~1. In this paper we focus on Obs.~1, 2, 4, 5 
and particularly on the intervals  that we identified as belonging to classes $\nu$, $\lambda$, 
$\chi$, and $\beta$ respectively \correc{\citet{belloni00}. To summarize, classes $\nu$, $\lambda$, 
and $\beta$ show  occurrences of spectrally hard X-ray dips of several hundred of second longs 
ended by a very bright X-ray spike, marking the return to a softer state (State-A).} \correc{These are followed 
by periods of intense and very variable X-ray activity indicative of State-A to State-B transition. 
We refer to these intervals (hard dips to soft spike) as cycles. Fig.~\ref{fig:Zoom246} shows zooms 
on cycles from each of the three observations. During class $\chi$ \grs\  has a relatively steady 
level of X-ray emission, over long times. It is in that case in the hard-intermediate state, and 
usually shows rapid ($>$1~s) variability and presence of Low Frequency QPOs}.

\subsection{\integral\ data reduction}
\correc{The data were reduced with the {\tt{Off Line 
Scientific Analysis (OSA)}} v7.0 software. Our first step was to produce ISGRI images 
in two energy bands, 20--40 and 40--80 keV, in order to detect the active sources 
of the FOV which have to be taken into account in the (spectral and temporal) 
extraction processes.  This list of ``bright'' sources was then given as an 
input for the extraction of IBIS/ISGRI spectra and light curves. This is a necessary
process} since all bright sources in the field of view can be considered 
as contributing to the background for the others. This means that we extracted spectra 
\correc{ and light curves} from between 5 and 8 sources. In the case of JEM-X the spectra were extracted 
from unit 1 of the detector\footnote{JEM-X unit 2 has been switched off since March 08, 2004}. 
For both detectors (IBIS and JEM-X) the latest versions (\correc{provided with {\tt{OSA 7.0}}}) 
of the response matrices were used in the spectral fitting. 
Note that due to the high variability of \grs\
accumulating spectra over the entire observation is meaningless most of the time. 
In the four observations we discuss here, we extracted time-resolved 
spectra based on selected good times intervals (gti). \correc{The JEM-X light curves 
were extracted with a time resolution of 1s between $\sim3$--$13$ keV, and 
further rebinned to 20~s. The ISGRI light curves were 
extracted between 18--50~keV with a time resolution of 20~s, and further rebinned to 100 or 200~s}.\\
\indent We should remark here that since \grs\ is very bright in the soft X-rays 
good quality spectra 
are obtained even with short accumulation times. This is, however, not the case above 20~keV.
We therefore used different binning of the redistribution matrix file (rmf) during 
the extraction process to obtain the spectra of the best possible quality.\\ 
The resultant spectra were fitted in {\tt XSPEC} v.11.3.2t, between 5 and 20~keV for 
JEM-X\footnote{Except in the case of Obs.~4 where the JEM-X spectra were fitted between 3 and 20~keV since 
comparison with a simultaneous PCA spectrum showed a good agreement between both detectors.}, 
and 18 and 300~keV for ISGRI. In most cases, the spectra were further rebinned 
so as to obtain good 
quality spectra (especially in the case of short accumulation times). We added 
3 \% systematics to the uncertainties in the JEM-X spectra, and 2\% to the ISGRI ones
 before the fitting process. In all fits a normalization constant was added to account 
for uncertainties in the cross calibration of the instruments. If frozen to 1 for the ISGRI 
spectrum, it is usually found to be between 0.9 and 1.1 for JEM-X.
Since  SPI is less sensitive than ISGRI below 200~keV we do not analyze 
SPI data in the study presented here. 

\subsection{\rxte\ data reduction}
The \rxte\ data were reduced with {\tt LHEASOFT} v. 6.1.2. following standard criteria 
for the definition of good times intervals \correc{(elevation above Earth limb greater 
than $10^\circ$ and offset pointing less than $0.02^\circ$)}. We first produced light curves 
in several energy ranges to produce color-color diagrams, hardness ratio, and 
hardness intensity diagrams (HID). For the latter purpose, we extracted 16~s light curves from 
the top layer of PCU~2 in four energy ranges, namely  2--126~keV (channels 0--255),  
band~1 = 3.28--6.12~keV (channels 8--14), band~2 = 6.12--10.22~keV (channels 15--24), and 
band~3 = 14.76--40.41~keV (channels 36--95). These light curves were background corrected 
before producing hardness ratio (or colors) defined as follows: HR1 = band~2/band~1 
and HR2 = band~3/band~1. Note that the background files were produced with the latest 
version of the saa-history file, release after the discovery of gaps in the former file.\\
\indent In the case of Obs.~4 (the steady state observation; Paper 1), we  
extracted PCA and HEXTE spectra from the unique interval that was simultaneous with 
\integral\ observation and observation from the RT. The procedure is the same as 
that described in \citet{rod03}, again apart from the newer version of the reduction 
software, and the background maps used here, that are the latest made available by the 
XTE SOF.\\
\indent For the production of high temporal resolution light curves and Fourier 
analysis of the PCA data, 
the data were reduced in the same way as presented in  \citet{rod04} (except for 
the different versions of the software). We extracted $\sim 4$~ms resolution 
light curves from {\tt{Binned}} 
data format covering 
the 2--15~keV energy range and $\sim 1$~ms light curves from {\tt{Single Bit}} and {\tt{Event}} data format 
respectively covering the $\sim5$--$15$~keV and $15$--$60$~keV energy ranges (we restricted the highest  
absolute spectral channel to 150 for the latter). We then produced power density 
spectra from the combined 2--60~keV (Nyquist frequency 
of 128~Hz) and 5--60~keV light curves (Nyquist frequency of 512~Hz)  
 to study the presence of low and high frequency quasiperiodic oscillations (LFQPO, HFQPO). 
Note that for the production of dynamical PDS, or PDS from a large energy range, the background correction 
can be neglected for the estimate of the fractional RMS variability. \\
\indent For Obs. 4 we also produced energy dependent PDSs in a manner similar to 
that presented in \citet{rod04}, i.e. extracting $\sim 4$~ms light curve from about
 20 spectral channels covering the 2--40~keV range. 
In that case the RMS amplitudes have to be corrected for the energy dependent background level.
This has been done according to \citet{berger94}, i.e. 
RMS$_{corr}=\mathrm{RMS}_{\mathrm{leahy}}\times\frac{S+B}{S}$
where $\mathrm{RMS}_{\mathrm{leahy}}$ is the RMS obtained from the Leahy normalized PDS 
(non background corrected), 
$S+B$ is the raw count rate, and $S$ the net count rate of the source.

\section{Results of the high energy observations}
We focus here on the observations for which we obtained simultaneous X-ray 
and radio observations (Paper 1). Amongst them, we distinguish Obs. 1, 2 and 5 
for which we observed sequences of X-ray dip/spikes (X-ray cycles) followed 
by radio flares, and Obs. 4 during which \grs\ is found in the steady $\chi$ 
class, and shows a relatively steady level of radio emission around 
40--80~mJy (Paper 1). \\

\subsection{Time-resolved spectroscopy and models for the spectral fittings of the cycles}
\begin{figure*}
\caption{Zoom on one cycle from each of the three observations with cycles. 
From top to bottom, the panels respectively represent the Ryle 15GHz, 
the JEM-X 3--13~keV, the ISGRI 18--50~keV lightcurves and 
the 3--13~keV/18--50~keV HR.   
 The different intervals from which spectra were 
extracted are indicated with letter on the JEM-X light curve.}
\plotone{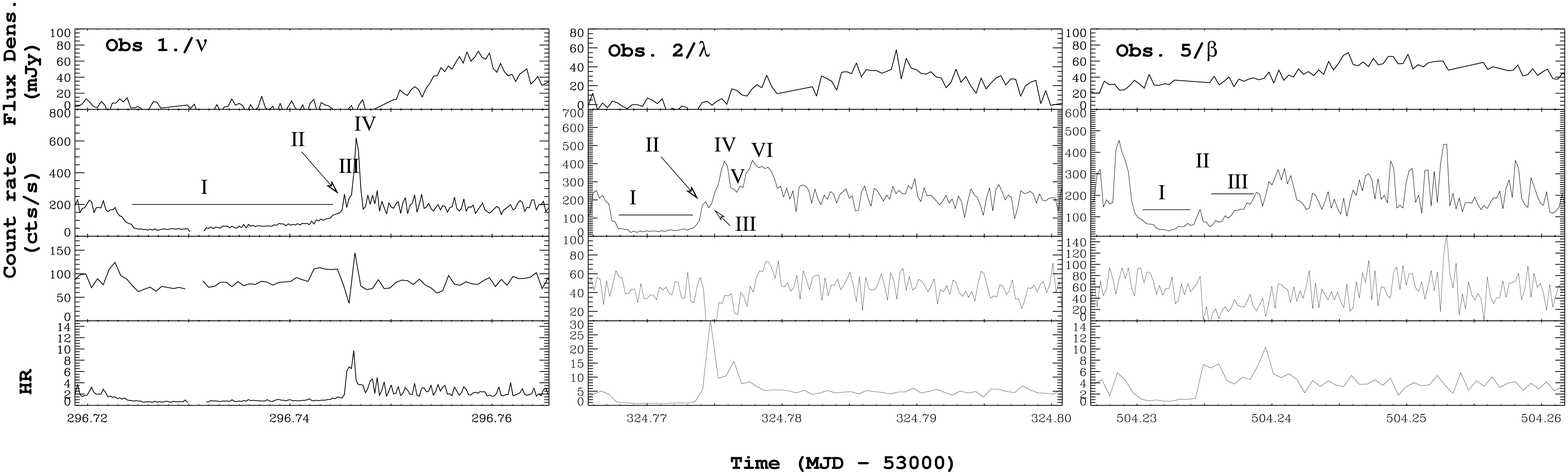}
\label{fig:Zoom246}
\end{figure*}

\begin{figure}
\epsscale{0.5}
\caption{Joint JEM-X and ISGRI spectra from the four intervals of Obs. 1/class $\nu$. The best-fit 
model is superposed onto the spectra, as well as the individual components, a disk and a Comptonized 
tail. The best fit parameters are given in Table~\ref{tab:fits}.}
\plotone{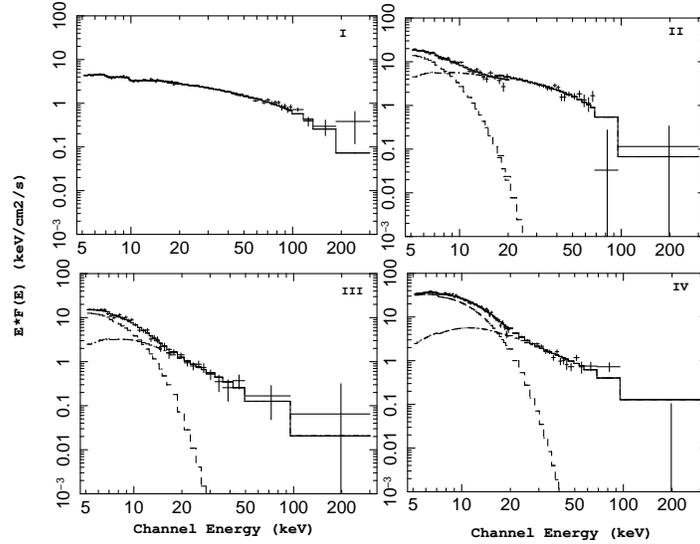}
\label{fig:nuspec}
\end{figure}

\indent All cycles from all classes show similar repeating patterns although their 
time constants are different. For example class $\lambda$ looks very much like 
class $\beta$, except that the sequences in the cycle seem to occur faster 
\citep{belloni00}. We can distinguish, in each cycle, at least three 
intervals (Fig.~\ref{fig:Zoom246}) with common 
characteristics (paper~1). Interval I is the spectrally hard X-ray dip, interval II 
corresponds to a precursor spike ending the dip, and interval III is a spectrally 
very soft dip immediately following the precursor spike.  In the case of class 
$\nu$ we identified 
an additional interval (Int.~IV) corresponding to the major spike following the whole sequence
(Fig.~\ref{fig:Zoom246}, left), while in class $\lambda$ three more intervals have been 
identified (Fig.~\ref{fig:Zoom246}, middle). A  major soft X-ray spike of about $600$~cts/s (Int.~IV), 
followed by a dip in both soft and hard X-rays (Int.~V), 
and a return to a high degree of soft and hard X-ray emission (Int.~VI). 
Since ejections seem to always occur after such cycles during classes $\nu$ and $\beta$ 
 \citep[e.g.][]{klein02}, we extracted time-resolved spectra from all
cycles, and further averaged those belonging to the same type of interval 
in those two observations. 
On the other hand it is the first time a radio flare is observed in class $\lambda$ (Paper 1). 
Since we cannot, a priori, know whether radio flares occur in response to each of these 
cycles,  we extracted time-resolved spectra from this particular cycle only. Note 
that the ubiquity of the accretion-ejection links during cycles seen in all other classes, 
indicates that each $\lambda$ cycle is probably followed by an ejection.\\
\indent During the spectral fittings, we tested several models, starting with the 
phenomenological family based on a power law tail at high energy plus a thermal component 
at low energy modified by interstellar absorption. Note that in all fits,  the 
absorption column density, $N_{\mathrm{H}}$, was frozen to $5.7\times10^{22}$ cm$^{-2}$ 
\correc{a value well within the 5--6$\times10^{22}$ cm$^{-2}$ range of value usually found in
the literature \citep[e.g.][]{belloni00}. Note that varying $N_\mathrm{H}$ within this range 
does not affect  the results of the spectral analysis significantly. This is especially true 
when considering the spectra above 5 keV}.
Because of a more accurate treatment of the inner regions of the disk, we used the 
{\tt{ezdiskbb}} model \citep{zimmerman05}, which assumes zero torque at the inner 
boundary of the disk. For Obs. 2
the statistics are not good enough to find other spectral features, or fit the 
data with more sophisticated models. 
A high energy cut-off is needed in intervals I$_\nu$ and I$_\beta$. 
Since a power law with an exponential cut-off is usually interpreted as due to 
thermal Comptonization, we then fitted all spectra of classes $\nu$ and $\beta$ with
a model consisting of interstellar absorption, a thermal component ({\tt{ezdiskbb}} when 
needed) and a Comptonized component ({\tt{comptt}}, \citet{titar94}). The temperature of 
the seed photons for Comptonization was set equal in 
all cases to that of the inner disk (except when the disk is not needed, and was then 
frozen to 0.7~keV, \correc{the temperature below which no disk component can be detected in 
JEM-X above 5 keV}).  In some cases the optical depth of the Comptonized component 
tends to very low 
values, while the electron temperature is quite high and poorly constrained. In this
ranges of parameters  we are outside the range of validity for this particular model. They correspond 
to spectral-states A and B \citep[in the classification by][]{belloni00} equivalent 
to steep power law/soft intermediate states, in which the coronal medium is very tenuous, 
and the population of electrons is probably non-thermal. 
The shape of the spectrum is then equivalent to a power law with no cut-off. 
In those cases the $\tau$ parameter was frozen to 0.01, \correc{which is the value towards what usually
tended the parameters, the smallest value allowed in {\tt{XSPEC}}} .
Note that although these caveats concerning the model used for the fitting, 
we preferred it to the phenomenological one because it 1) avoids
 the mixing of the two components at low energy (ie no overstimate of the contribution of 
the corona below 3kT$_{seed}$) that is unavoidable with 
a simple power law model. 2) we are interested in the relative 
variations of the 3--50~keV flux thus well bellow the supposed kT$_{e}$, in other 
words where the shape of the spectrum is power law-like. 3) The use of the same model 
allows a true and robust comparison of the fluxes in the different intervals. 
Although when $\tau$ is very low and kT$_{e}$ very high the interpretation 
in terms of thermal Comptonization is thus dubious, in terms of fluxes the relative 
contributions of each model component to the total flux are, therefore, 
more reliable. Note that similar evolutions of the fluxes are seen
when fitting with the phenomenological model. This shows our method is valid.
The best-fit parameters 
for all intervals of the three observations are reported 
in Table \ref{tab:fits}, while  Fig. \ref{fig:nuspec} shows the particular example of 
the four spectra from Obs.1/Class $\nu$ with the best model and individual 
spectral components superposed.\\
\indent  In all three observations the disk temperature increased smoothly through 
the cycle, with a simultaneous decrease of its inner radius (Table \ref{tab:fits}). 
Its overall contribution to the total 3--50~keV flux also 
increased accordingly. On the other hand the 3--50~keV contribution of the 
Comptonized component was never so monotonic. 
In class $\nu$ it first increased from I$_\nu$ 
to II$_\nu$, and was then reduced by a factor of $2.2$ from II$_\nu$ to III$_\nu$. 
It recovered at a higher level (a factor of approximately $1.7$ higher) in 
IV$_\nu$. A very similar evolution is seen in Obs.~5/Class $\beta$ where the 
Comptonized component sees its 3--50~keV flux reduced by a factor 6.7 between II$_\beta$ 
III$_\beta$. During Obs.~2/class $\lambda$ the disk appeared during III$_\lambda$ 
and remained at a high temperature and low radius over the remaining intervals. The 
flux of the hard tail (assumed to be the emission of the Comptonizing corona) first 
increased from I$_\lambda$ to II$_\lambda$, with a small decrease 
in the photon index. It was not statistically required 
in the fit to III$_\lambda$, which likely indicates that it had disappeared. 
We, nevertheless, added a power law to the disk component with a photon index 
frozen to 2.0\footnote{This value is a quite standard value for BH in general. Note 
that freezing the photon index to values greater than 2 results in even lower limits
on the flux of the hard tail component.} 
and estimated the upper limit on its flux. We find that from II$_\lambda$ 
to III$_\lambda$ the power law contribution to the 3--50~keV flux was  suppressed 
by a factor greater than 4.2 (at $3$--$\sigma$), and remained  at a low 
flux until VI$_\lambda$ where it reached its highest flux over the entire 
cycle (Table \ref{tab:fits}).

\subsection{Temporal Analysis of the cycle}
\begin{figure*}
\epsscale{1}
\caption{Dynamical power spectrum (top) and PCA light curve (bottom) from the three observations 
with cycles. a) $\nu$-type cycle simultaneously observed with \rxte\ and \integral. b)
$\lambda$-type cycles 
observed with \rxte. c) $\beta$-type cycle observed with \rxte. \correc{The gray scale represents the RMS power
of the variability, with white having a higher power than dark points.}}
\plotone{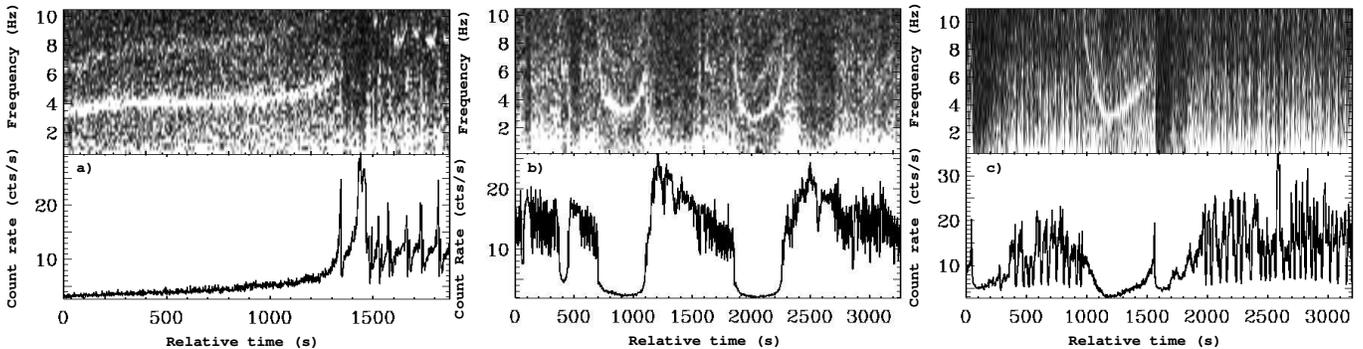}
\label{fig:dynpo}
\end{figure*}

\indent The dynamical power spectra of all three observations showing 
X-ray cycles are shown in Fig.~\ref{fig:dynpo}. In the case of  
type $\nu$, this corresponds to a cycle that  is simultaneously covered with \integral\ and 
\rxte, and followed by a radio flare. Although the dip is not entirely 
covered by \rxte, a strong feature, a low frequency QPO (LFQPO), of variable 
frequency can be seen. The frequency of the LFQPO seems to follow 
the evolution of the X-ray flux, as observed in other classes 
\citep[e.g.][]{markwardt99,rod02a,rod02b}, namely $\beta$ and $\alpha$. \\
\indent Very similar behavior is seen during $\lambda$ and $\beta$-type cycles: 
a strong LFQPO of variable frequency appears at the moment \grs\ enters the dip 
(Fig. \ref{fig:dynpo}). 
The frequency rapidly decreases, in (cor)relation with the X-ray flux. The feature 
ceases at the end of the dip. Because of the short 
time scales of the events occurring at the end of the dip in class $\lambda$ 
and the 16-s step on which the dynamical PDS is calculated, the precise moment 
of the cessation of the LFQPO is  quite difficult to estimate. It seems to occur somewhere 
between the equivalent of Int. B to D (Fig. \ref{fig:dynpo}). In class $\beta$
the QPO is cut at the maximum of the precursor spike, and so 
seems to be the overall noise (Fig. \ref{fig:dynpo}). Note that for the two 
latter cycles, we do not have simultaneous \rxte\ coverage
 to the cycles which were followed by the radio flares. We can, hence,  
hypothesize that \grs\ undergoes the same behavior 
concerning the LFQPO during all cycles, and thus also showed a LFQPO prior 
to the ejections during classes $\lambda$ and $\beta$. Given that this is commonly
observed in \grs, and that it has already been seen in class $\beta$ 
\citep[e.g.][]{markwardt99,rod02a} in simultaneity with radio flares \citep{mirabel98}, 
this hypothesis seems quite reasonable.\\

\subsection{Observations 4: Class $\chi$}
\subsubsection{X-ray spectral analysis}
\begin{figure}
\epsscale{0.5}
\caption{Zoom on the interval of Obs. 4. from which we extracted the time dependent spectra.
From top to bottom, the panels respectively represent the Ryle 15GHz, 
the JEM-X 3--13~keV, the ISGRI 18--50~keV, and PCA 2--60~keV lightcurves.  
The different intervals from which spectra were extracted are indicated with the 
letter on the light curves. The best fit parameters are given in Table~\ref{tab:fitschi}.}
\plotone{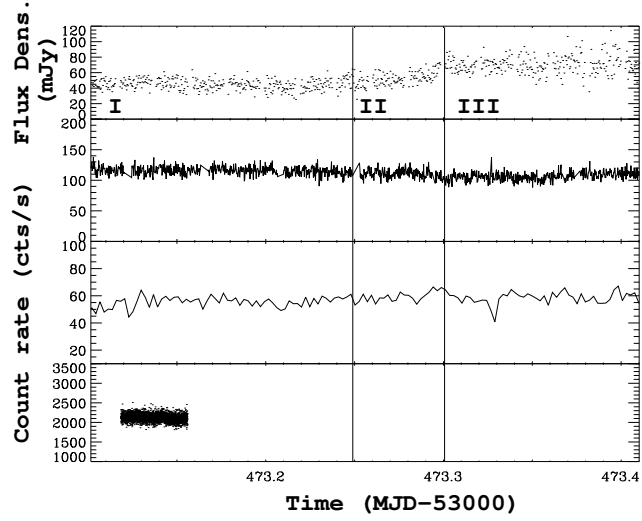}
\label{fig:zoomchi}
\end{figure}

\begin{figure}
\epsscale{0.75}
\caption{{\bf{Left:}} Residuals in terms of $\sigma$ between the models used to fit the \rxte\ spectra.
In each case the model mentioned in the panel is convolved with interstellar absorption. Pl stands for power law 
and ga stands for Gaussian. \correc{From top to bottom the reduced $\chi^2$ are the following: 25.2 (132 d.o.f.), 
7.3 (131 d.o.f.), 5.5 (129 d.o.f.) and 1.8 (126 d.o.f.)}. {\bf{Right:}} Joint
\integral+\rxte\ spectra from Obs.~4 Int.~I. The best model (\correc{consisting of a {{\tt comptt}} a power
law and a Gaussian convolved by absorption}) is superposed on
the spectra.}
\plotone{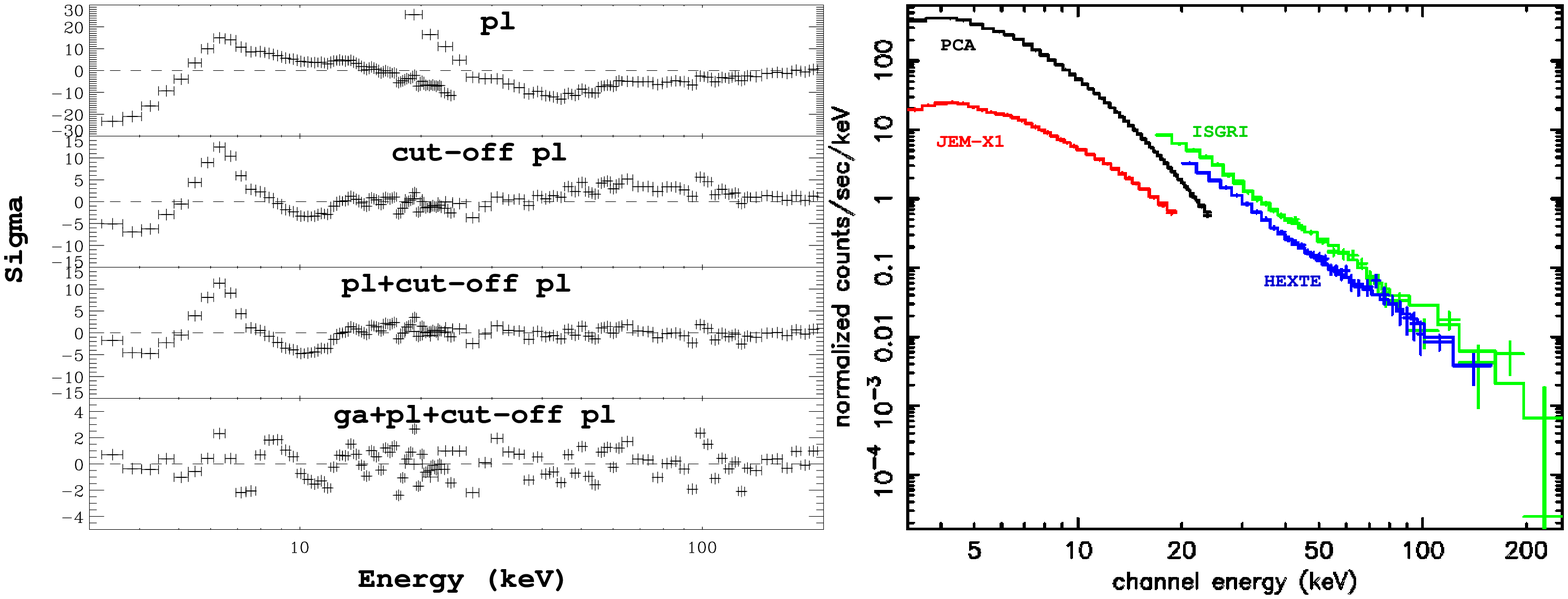}
\label{fig:residuals}
\end{figure}

During the \integral\ coverage of Obs. 4, we obtained two observing intervals 
with the RT and five with \rxte\ (Paper 1). The second coverage with the RT 
overlapped only very briefly with \integral\ and we will not discuss it further here.
During the first, however, the multiwavelength coverage (Fig. \ref{fig:zoomchi}) shows that 
the radio flux,  while being rather steady during the first 
 $13$~ks at a mean value of $44.9$~mJy, increased slightly between MJD 53473.25 and 53473.30. 
It reached a mean level of $70.4$~mJy after 53473.30 up to the end of the radio coverage
 on MJD 53474.42. We separated this observation 
into 3 intervals (I$_\chi$, II$_\chi$, III$_\chi$) based on the radio behavior of 
the source from which we extracted \integral\ spectra. In the case of I$_\chi$ 
an \rxte\ pointing is also available and was added to the spectral fit. \\
\indent As in the previous observations, we started with different combinations of 
phenomenological models. 
It has to be noted that none of the spectra is well represented by the standard model 
composed of an accretion disk and a power law tail. A single absorbed power law is a very poor 
representation of the spectra. Fig. \ref{fig:residuals} shows the residuals in terms of 
number of $\sigma$ between the model and the spectra. Adding a multicolor disk black 
body in either of its forms ({\tt{diskbb, ezdiskbb}}) \correc{still leads to a poor 
representation of the spectrum (even with a Gaussian, see below).  In I$_\chi$, for 
example, the reduced $\chi^2$ is 10.4 for 124 d.o.f.}. Since the simple power law 
overestimates the spectrum at high energies, we replaced it by a cut-off power law. This provides
a much better representation of the spectrum, although still not satisfactory. \correc{Again 
adding a disk emission does not provide a good description of the spectrum. In I$_\chi$, for 
example, even after having included a Gaussian around 6.5~keV, the reduced $\chi^2$ is 
 12.6 for 123 d.o.f.}.  The emission 
above  $>40$~keV is still underestimated. \correc{Instead of a disk emission, }an extra power 
law improves the fit significantly (F-TEST probability of $\sim7\times 10^{-19}$ chance improvement). 
Some residuals are still visible in the iron K$_{\alpha}$ region at $6.5$~keV. We therefore 
added a Gaussian to account for this feature. \correc{The reduced $\chi^2$ is 1.73 (123 d.o.f.) 
in I$_\chi$ with this model.}. 
The cut-off power law was then replaced  by the {\tt{comptt}} model \citep{titar94}. This 
model provided very good fits to the data \correc{and gives a better $\chi^2$  (Table \ref{tab:fitschi})}. 
All spectral parameters are reported 
in Table \ref{tab:fitschi}.\\
\indent A clear evolution in terms of spectral parameters is seen between the 
three intervals of Obs.~4 (Table \ref{tab:fitschi}), especially for the power law 
component.  The temperature of the seed photons for Comptonization is rather stable around 
0.3~keV, and the Kompaanets parameter 
($y=kT_e/m_ec^2\times {\mathrm{max}}(\tau,\tau^2)$), that roughly characterizes 
the efficiency of Comptonization, is multiplied by a factor of about $1.1$ between 
I$_\chi$ and III$_\chi$. In the mean time the power law photon index of the hard tail 
gets softer. In terms of fluxes (unabsorbed 2--200~keV), that of the Comptonized 
component is divided by a factor of about 1.2 between I$_\chi$ and III$_\chi$, while that 
of the hard tail is multiplied by a factor of about $1.6$. The power law fluxes are significant 
at more than 10-$\sigma$ in each intervals. Based on the errors of the power law 
parameters we can estimate a ratio 1.61$\pm0.13$ (1-$\sigma$ error) 
increase \correc{of the power law flux} between I$_\chi$ and  III$_\chi$.

\subsubsection{Timing Analysis: LFQPOs}
\begin{figure}
\epsscale{0.5}
\caption{Spectrum of the LFQPO from the unique \rxte\ observation that was 
simultaneous to \integral\ Obs. 4 and the RT. The best model obtained from fits with power law and power law 
with an exponential cut-off are respectively over-plotted as a dash-dotted and a dashed line. A better fit is obtained 
with the power law with cut-off.}
\plotone{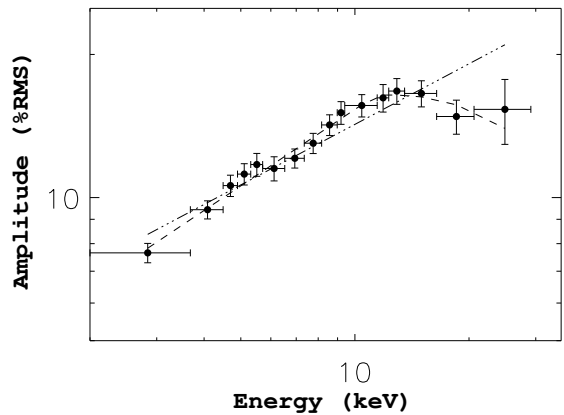}
\label{fig:specqpo}
\end{figure}

We first analyzed the 2--60~keV PDS of the unique  $\sim3000$~s \rxte\ pointing 
that was simultaneous with both \integral\ and the RT.  The PDS was computed between 
0.03125~Hz and 124~Hz. The PDS is composed of a flat top component up to a break 
frequency above which a power law decrease can be seen. On top of this continuum 
a strong Low Frequency Quasi Periodic Oscillation (LFQPO) is detected. Above about 
$20$~Hz the PDS is compatible with Poissonian noise. We modeled it with the sum 
of a broad zero-centered Lorentzian, a narrow one (the LFQPO), and a constant to 
account for the white noise. The fractional RMS amplitude of  the continuum of the PDS 
above 0.03125~Hz is $13.1\pm0.4\%$. The QPO centroid frequency is 
$3.77\pm0.01$~Hz, its quality factor Q($=\nu_{\mathrm{centroid}}/{\mathrm{FWHM}}$) is 
7.7, and its fractional RMS amplitude $10.9_{-0.7}^{+0.6}\%$.\\
\indent This best model was used to fit the energy dependent PDSs. Note that given their poorer
statistics and the fact that we are interested in the parameters of the LFQPO, we subtracted 
the white noise and restricted the frequency range to 0.03125--10 Hz. The energy dependence of 
the LFQPO (hereafter QPO spectrum) is plotted in Fig. \ref{fig:specqpo}. The QPO spectrum
is hard in the sense that it increases with energy. A plateau and possible following 
decrease can be seen from 10--15~keV. Such behavior has already been reported for 
LFQPOs from class $\chi$ observations \citep{rod04}.  We first fitted the QPO spectrum 
with a simple power law model. The statistics of the fit is poor with $\chi^2_\nu$=2.05
for 14 d.o.f.. In order to account for the deviation from a simple power law 
 (Fig.~\ref{fig:specqpo}), we added an exponential cut-off to the power law  
(see \citet{rod04}). We obtained a good 
representation of the spectrum with $\chi^2_\nu$=0.41 for 12 d.o.f.. The best parameters are 
a power law index of $0.53\pm0.04$, a cut-off energy $E_{\mathrm{cut}}=11.3\pm1.3$~keV, 
and a folding energy for the cut-off $E_{\mathrm{fold}}=23\pm6$~keV. 

\section{Discussion}
\subsection{Model-independent evolution, hardness intensity diagrams}
\begin{figure}
\epsscale{0.5}
\caption{Hardness-Intensity diagrams of the four observations discussed in the paper.
Each point represents 16s. {\bf{Left:}} PCU 2 count rate vs. (6.1--10.2)~keV/(3.3--6.1)~keV.
 {\bf{Right:}} PCU 2 count rate vs. (14.8--40.4)~keV/(3.3--6.1)~keV. The dashed line is roughly indicative 
 of the position of the jet line: each time \grs\ transits from the dip through this line an 
 ejection occurs.}
\plotone{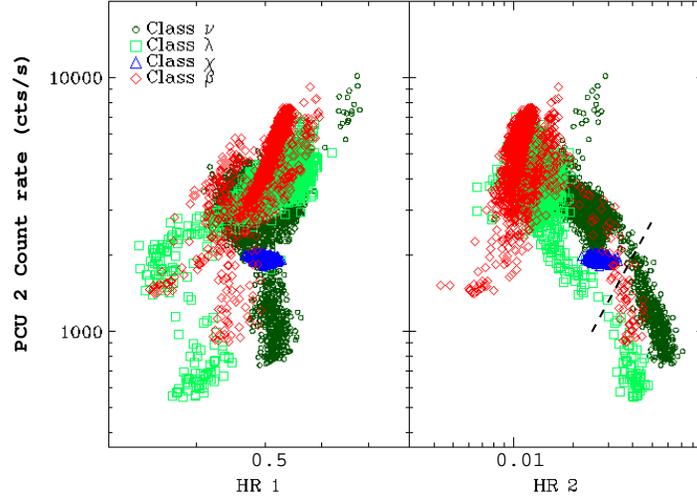}
\label{fig:hid}
\end{figure}

Fig. \ref{fig:hid} shows two  HIDs 
of all the \rxte\ pointings of the four observations presented in this 
paper (see Table 2 of Paper 1 for the journal of the entire observations).  
The first one, representing the PCU2 intensity vs. HR1 
(\correc{(6.12--10.22)~keV/(3.28--6.12)~keV})
(hereafter HID1) matches exactly the colors commonly used in the 
literature for the production of HIDs \citep[e.g.][]{belloni05,fender04}.  
However (as clearly mentioned in \citet{fender04}) since the thermal 
disk component in \grs\ is always brighter than in any other BH, using the 
same colors results in having the hard C-state lying in a softer part of the 
HID than the soft B-state (Fig. \ref{fig:hid}, left). The second HID (HID2) 
makes use of colors \correc{(HR2=(14.76--40.41)~keV/(3.28--6.12)~keV)} well 
separated so that the disk component can clearly be separated from any other component. 
Although the values of the HR are quite different between
the three classes with cycles, the general pattern shows striking
similarities in both plots. They all have in common a ``parallel'' trend
from the low values of intensity and HR1$\sim0.5$, (resp. HR2$\sim0.03$)
to intermediate values of the intensity, and HR1$\sim0.5$, 
(resp. HR2$\sim0.01$). This corresponds to the dips of the cycle (at the 
right bottom of the tracks), and the spikes that end them. This parallel
track also corresponds to the temporal evolution of \grs\ when going 
through the cycle. It first evolves in intensity at a high value of HR2 (or 
roughly constant value of HR1), and, after the spike,  falls to the left
part of the plots (low values of HR2 and decrease in intensity). 
Note that, as we show in the next section, the spike is very likely the trigger of 
the ejection, hence the HR2 of the spike could represent the jet 
line of \citet{fender04}. Its precise value is, however, quite difficult to estimate
in the three classes, as just before the ejection, the values of HR2 are clustered 
in a region they can reach even in the flaring state. This is quite well illustrated 
in Fig. \ref{fig:hid} in the case of class $\lambda$ around HR2=0.02. 
However we remark that each time \grs\ exits the dip of a cycle, an ejection occurs 
(Paper 1).  The approximate position of the jet line is represented in the right
panel of Fig. \ref{fig:hid}. The flaring states of the observations correspond to 
oscillations between the points that lie on the left of this jet line. 

\subsection{Ejection of coronal material?}
In Paper 1 we have shown that the same sequence of events will lead to 
the ejection of material. Namely, an X-ray dip of duration longer than 100~s 
terminated by an X-ray spike always precede a radio flare. We have argued, in
a model-independent way, that the X-ray spike terminating the sequence was 
the trigger of the ejection. This was based on the similarity of the sequence of 
events, and that in class $\beta$ the spike had been identified by \citet{mirabel98}
as the trigger of the ejection. Here we followed a different approach, and 
performed fits to each remarkable interval of the cycles.\\
\indent Although the precise spectral parameters returned from the fits 
(Table \ref{tab:fits}) show differences between the cycle of each class, a 
similar evolution can be seen in all. As presented in Sec. 3, while the evolution
of the disk along the cycles is monotonic, and may indicate a disk that slowly 
reaches its last stable orbit around the BH, the evolution of the Comptonized component 
or the power law component (the corona) is quite different. While it entirely dominates 
the spectra during the hard-dips (Ints.~I), it disappears after the spike labeled 
II in all cases, and seems absent in the soft dips (Ints.~III).
Its relative contribution to the source flux, in the same band as that of the disk, 
decreases by a minimum factor of 2.2 (in class $\nu$).\\
\indent In the case of class $\beta$ \citet{mirabel98} demonstrated that 
the X-ray spike was the trigger of the ejection. In Paper 1, based on the similarities of 
the sequences seen during the three classes, shape, evolution of the source fluxes 
in different X-ray bands, and delays between the X-rays and the radio peak, we argued that 
the  spike (II$_\nu$, II$_\lambda$ and II$_\beta$) is the very likely trigger of the 
ejection. Our spectral analysis is compatible with such an interpretation, since in all cases
the spectra greatly evolve after the spike, and show a transition to a state dominated 
by the disk component. In that case, the quantitative evolution of the coronal flux 
shows that this component disappears between Ints.~II and III. Given the detection of 
ejected material at approximately the same time delay in all three classes, it is tempting
to consider that the corona is ejected at the spike, decoupled from the central system, and is 
further detected at radio wavelengths.\\
\indent One could argue that the spike is not the real trigger. The other remarkable 
moment of each cycle (at least from the spectral point of view) is the entrance into 
the dip. Here the spectrum becomes hard, which likely indicates that the inner 
regions of the disk either recede from the BH or disappear.
Several arguments do not favor this moment as being the trigger of the ejection. 
First the radio flares --the signature of the ejections-- have similar profiles 
(Paper 1), while the transitions to 
the dip at X-ray energies are quite different, especially in their temporal evolutions. 
The decrease of the X-ray count rate is quite slow in $\nu$ and $\beta$, while it 
is quite sudden in $\lambda$. If this moment were the signature of the ejection 
at X-ray energies, it is reasonable to think that its temporal evolution would 
influence the subsequent ejection. Secondly, during class $\beta$, a 0.8~keV disk 
is clearly detected during the dip, while the amplitude of the ejection is 
higher than in class $\lambda$ (no disk in the dip), and the duration of the 
dip is longer. This is in contradiction with a model where the longer 
duration dip associated with a higher amplitude ejection would indicate that 
more disk material is ejected, and 
hence it would take longer for it to be refilled. Finally the delay between the peak 
in radio and the spike terminating the cycle is similar in all classes (Paper 1), 
while the duration of the dip, and hence the delay between the transition into the 
dip and the radio peak is quite different between the three classes. It is around 
$0.8$~hour in class $\nu$, while it stays around $0.4$~hour in the other two classes. \\
\indent We therefore favor an interpretation where the coronal material 
is ejected from the system at the spike terminating the hard dips of the cycle.
It is interesting to note that in another class with a cycle (class $\alpha$), 
\citet{rod02b} had come to the same conclusion, although based on the inspection of 
the evolution of HRs and light curves only (see also \citet{vadawale03}). 
Further strengthening this interpretation is the case
of another microquasar, XTE~J1550$-$546, where \citet{rod03} 
arrived at the same conclusion studing the evolution of this source's outburst in 2000.
 The similar conclusions reached in two different sources may 
indeed indicate that the discrete ejections in (all) microquasars are composed of 
coronal material.

\subsection{Steady hard (intermediate) state}
\begin{figure}
\epsscale{0.5}
\caption{{\bf{Top panel:}} \rxte/PCA spectrum of Obs. 4 Int. A (bullet). The best fit model, the 
power law component, and the {\tt{comptt}} component are over plotted. {\bf{Bottom panel:}} Relative
contribution of the {\tt{comptt}} component and spectrum of the QPO.}
\plotone{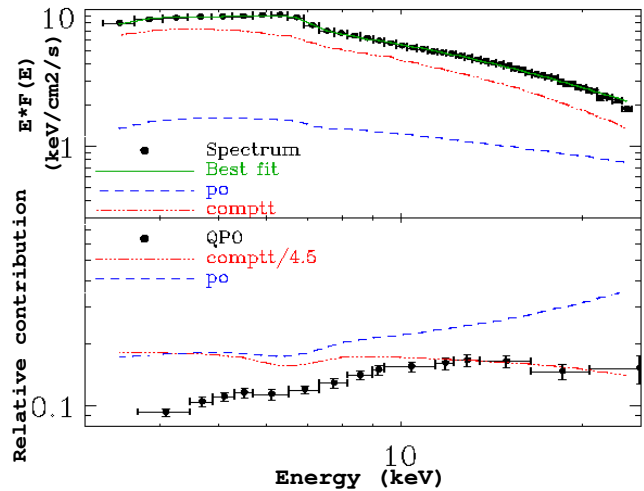}
\label{fig:contrib}
\end{figure}

During one of our observations, \grs\ was found in the so-called $\chi$ class
of variability in the X-rays, which corresponds to its hardest steady state\footnote{Note that 
as \grs\ is never observed to be in the canonical hard state seen in other microquasars, 
``hard'' is taken here as relative to its other spectral states. The so-called
$\chi$ class/state corresponds to the hard-intermediate state in the other sources,  
following the definition of \citet{homan06}}.
In the radio domain a relatively  bright and steady emission is detected, and 
attributed to the presence of a compact jet \citep[e.g.][]{fuchs03}. As mentioned 
in \cite{vadawale03} the spectrum of this radio loud $\chi$ observation indicates
the presence of a Comptonizing component and an extra power law (Table \ref{tab:fitschi}). 
We do not need any thermal component at low energy, which indicates that the accretion 
disk is quite cold (at least it has no effect above 3~keV), and indeed the temperature of 
the seed photons for Comptonization is quite low, around 0.3~keV. It is interesting to note
that when separating the observation into three noticeable intervals based on differences
in the radio fluxes, we clearly see the spectral evolutions of the two X-ray 
spectral components (Table \ref{tab:fitschi}). While the 15 GHz flux increases,
 the 2--200~keV flux of the Compton component decreases, while that of the 
extra power law component seems to increase as well in coordination with a softening of 
the tail. \\
\indent The spectral characteristics of radio-loud $\chi$ classes in \grs\
have been shown to always imply Comptonization by a corona of a few keV and optical
depth of the order of several units \citep{trudolyubov01,vadawale03}.  This  either 
results in spectra of shape similar to the ones we observed here \citep{vadawale03},
 or phenomenologically with breaks at 12--20~keV \citep{trudolyubov01}. Several 
competing models may explain the simultaneous presence of thermal Comptonization 
and of an additional power law. This may be the signature of Comptonization on
a hybrid thermal/non-thermal population of electrons \citep[e.g.][]{pouta96}. This 
family of model has been successfully applied to the modeling of microquasars in 
steep power-law/soft intermediate/very-high states showing power laws with no 
cut-off up to the MeV domain \citep[e.g.]{cadolle06}, or even to some peculiar classes
of \grs\ \citep{hanni05}. These states were, however, never associated to a strong radio
emission. \citet{trudolyubov01}
 argued that differences between the radio loud and radio quiet $\chi$ states
 are related to different structures of the accretion flow around the black hole,
 the obvious difference being the presence or absence of a compact jet.
The fact that the flux of the power law component increased in simultaneity with the radio
flux from the compact jet seems to indicate a link between the hard tail and the 
radio emission. This correlation between the evolution of both components leads 
to the tempting conclusion that the hard tail seen at X-ray energies is emitted by 
the jet itself. 
This possibility may be further re-inforced by the very similar evolution of both components: 
the 2-200~keV flux of the hard tail 
increased by a factor 1.61$\pm0.13$ between I$_\chi$ and III$_\chi$ when at the same time 
the radio flux increased by a factor 1.57$\pm0.14$. \correc{The Pearson correlation factor 
between the radio and power law fluxes of the three intervals is 0.93, which suggests that the 
radio and power law fluxes are strongly correlated}. In the case emission from the 
compact jet the radio emission would lie in the optically thick spectral domain, while the 
X-ray emission would lie in the optically thin one. Therefore a direct linear relation may therefore
appear as a special case.  It has to be noted that the parameters
of the power law we find here are very similar to the one reported by \citet{vadawale03}.
In the latter work, these authors suggested, based on a simple 
modeling of a compact jet, that this additional power law could be direct 
synchrotron emission from the jet. The similarity of their results and our 
may also favor the same interpretation.
The origin or possible link of the Comptonized component with the jet is less clear as 
inverse Comptonization can occur either on the base of the jet, or under the form of SSC of 
the jet photons \citep{markoff05}. The fact that the temperature for the seed photons 
remains constant, at a low value, rather suggest the seed photons for Comptonization 
probably come from the cold disk. \\
\indent As expected during  class $\chi$, a strong LFQPO is detected during 
the unique interval that is simultaneous to the RT and \integral. We confirm
previous findings that the QPO spectrum shows a cut-off \citep{rod04}, and by comparing 
with previous observations, we can see that the cut-off energy is clearly not constant 
and may evolve from one observation to another \citep{rod04}. In \citet{rod04}
we had suggested that the spectrum of the QPO could be understood if one assumed 
that the high energy component was not modulated and came from a jet. Here the same 
conclusion holds when observing a cut-off while the power law may extend to much higher 
energies than the QPO does (Fig. \ref{fig:specqpo} \& \ref{fig:residuals}). The presence 
of a LFQPO is always contemporaneous to the presence of a Comptonized component in the energy
spectra of the source. Whether this component is thermal or not (i.e. whether 
it shows a cut-off or not when modeled with power laws) depends on the source and the 
spectral state. Several models have attempted to explain the QPOs. One may distinguish 
between models based on instabilities in the disk 
\citep[e.g., the Accretion Ejection Instability (AEI)][]{tagger04}, oscillations 
of the boundary between the accretion disk and an inner post-shock region 
\citep[e.g.][]{chakrab95}, or even global oscillations of the disk itself  
\citep[e.g.][]{nowak92}. The correlations  seen between some of the spectral 
parameters and those of the LFQPO as well as their intrinsic properties 
\citep[e.g.][]{markwardt99,rod02a,rod02b,vignarca03,rod04} indicate that they are tightly
linked to the Keplerian orbit of the accretion disk, and to the corona itself.
In order to study the possibility that the QPO represents some oscillation of the 
corona, we plotted in Fig.~\ref{fig:contrib} the $E$--$F_E$ PCA spectrum from I$_\chi$
 with the different contributions superposed upon it. In the lower panel we compared
the relative contribution of the {\tt {comptt}} and power law components to the QPO spectrum. 
The QPO spectrum is clearly not power law like further indicating 
that its origin does not lie in the hard tail. The fact that this hard tail is likely 
the direct emission from the jet at high energy gives more strength to the suggestion of 
\citet{rod04} that the jet emission do not contain any quasi-periodic modulation. 
Furthermore the QPO spectrum and the relative contribution of the 
Comptonized component, even scaled down, are clearly not 
compatible (Fig.~\ref{fig:contrib}). This may rule out models based on the oscillation 
of the corona, and would rather favor models of disk instability 
\citep[as e.g. the AEI; ][]{tagger04}, 
or global oscillations of the disk \citep{nowak92}. 
The presence of strong QPO together with a strong corona may simply indicate that 
the corona either 
enhances the modulation, or simply that the QPO needs the corona to develop.
In that case one has to note that whatever produces the QPO,
its emission necessarily undergoes Compton up-scattering. 
 Note that if the Compton component comes from SSC from the jet, then the large 
differences between two spectra (QPO and Compton) further indicates that the jet is not 
at the origin of the LFQPO. 

\section{Summary}
We have presented the X-ray spectral analysis of four \integral\ and \rxte\ observations 
of the microquasar \grs. In this analysis we focused on intervals that were 
strictly simultaneous to radio observations taken with the Ryle Telescope.
The results of the multiwavelength approach are reported in Paper~1, where we 
showed that bubble ejections always follow a sequence of a hard dip terminated 
by a sudden spike at X-ray energies (referred to as a cycle).
We performed a fine spectral analysis of the entire cycles 
from a few keV up to about $100$~keV for the first time in three different classes. We then
presented a spectral and timing analysis of the unique observation during which 
\grs\ is found in a steady hard-intermediate state, and for which a strong and rather 
steady jet is detected in radio. The main results of our analysis can be summarized 
as follows:
\begin{itemize}
\item In all classes with cycles ($\nu$, $\lambda$ and $\beta$) we showed that the ejection 
seen in radio is triggered at the spike terminating the cycle. By comparison with 
other such classes which always end with a large X-ray spike (e.g. $\theta$ and $\alpha$),
we suggest that this is a very generic behavior in \grs.
\item In the same observations, we showed that each spike is indicative of the disappearance
of the Compton component from the X-ray spectra. We interpreted this as the evidence that 
the ejected material was the corona responsible for the Comptonized component. Again
comparing with the behavior seen during other classes \citep{rod02b,vadawale03} or in other 
sources \citep[e.g.][]{rod03} leads us to suggest that this phenomenon is very generic in 
microquasars.
\item In all cycles a strong LFQPO with variable frequency shows up at the transition into the
dip and is quenched at the X-ray spike ending the cycle. This phenomenon may indicate a strong
link between the QPO, the X-ray behavior, and the subsequent ejection. In a model-dependent
interpretation, this is compatible with the prediction of the AEI and the magnetic flood
proposed to explain the behavior of \grs\ during class $\beta$ \citep[Paper 1][]{tagger04}.
\item The X-ray behavior of \grs\ during a radio loud $\chi$ observation shows the presence
of two emitting media, one responsible for thermal Comptonization, and another emitting  a hard
X-ray tail modeled with a power law extending up to 200~keV without any break.
\item We find a  correlation between the radio flux and the 2--200~keV flux of the
hard tail. The very similar evolution of both the radio flux, and the flux of the hard tail
 may suggest that this component is direct  emission from the jet.
\item The energy dependence of a LFQPO does not follow the energy dependence of the relative
contribution of the Comptonized component, neither that of the hard tail. This indicates 
that the QPO does not have its origin in the jet, and may also rule out 
models based on oscillations of a Compton corona. Globally the behavior of the LFQPO is more 
compatible with models of disk instabilities.
\end{itemize}

All these points tend to show \grs\ behaves in a way very similar to other microquasars, and 
is certainly compatible with the generic model of \citet{fender04}. The main difference is 
that its disk is never completely emptied and hence \grs\ never reaches true quiescence 
and undergoes cycles between bright hard intermediate and soft intermediate states.
 The similarity of 
all cycles, the possible correlation between the amplitude of the ejections and the 
duration of the dip at X-ray energies (Paper 1), the repeating scenario of ejection of the 
coronal medium, the correlated presence of LFQPO indicate a unique mechanism may be 
at the origin of all cycles. In the case of a $\beta$ cycle \citet{tagger04} proposed 
a magnetic flood scenario compatible with these observed properties.  
Our monitoring campaign will continue with the view to cover many more of 
those cycles and better understand the connections and origin of the hard X-ray emitters to 
those of the radio jet

\begin{acknowledgements}
J.R. would like to thank A. Gros for invaluable help with the ISGRI data reduction software,
and  C.A. Oxborrow for precious help with the JEM-X data reduction and calibration. 
DCH gratefully acknowledges the Academy of Finland. AP acknowledges the Italian Space 
Agency financial and programmatic support via contract ASI/INAF I/023/05/0.\\
Based on observations with \integral, an ESA mission with instruments and science data centre funded by ESA member states (especially the PI countries: Denmark, France, Germany, Italy, Switzerland, Spain), Czech Republic and Poland, and with the participation of Russia and the USA.
This research has made use of data obtained through the High Energy 
Astrophysics Science Archive Center Online Service, provided by the NASA/
Goddard Space Flight Center.
\end{acknowledgements}

\bibliography{ms}

\newpage
\begin{table}
\caption{Best parameters obtained from the fits to the JEM-X+ISGRI data 
of the several intervals of Obs. 1, 2 and 5. The fits of Obs.~1 correspond to the 
interval represented in Fig.~\ref{fig:nuspec}. For Obs. 1 \& 5 the best fit model is 
{\tt{phabs(ezdiskbb+comptt)}} (in {\tt{XSPEC}} notation), while for Obs. 2 it is 
{\tt{phabs(ezdiskbb+powerlaw)}}. 
The errors are given at the 90$\%$ confidence 
level. The upper 
limits on the fluxes are at the $3$--$\sigma$ (99.73$\%$) level.}
\begin{tabular}{cccccccc}
\hline
\hline
Int. & kT$_{seed}$ & R$_{in}$\tablenotemark{a} & kT$_e$ & $\tau$ & $\chi_\nu^2$ & \multicolumn{2}{c}{unabs. fluxes }\\
     & (keV)       &  ($\times\frac{\mathrm{R}_\mathrm{G}}{\sqrt{cos(i)}}$ )& (keV) & & (d.o.f.) & 3-50~keV disk & 3-50~keV comptt\\ 
     &             &                           &        &        &              & \multicolumn{2}{c}{($\times10^{-8}$~\ergcms)}\\
\hline
\multicolumn{8}{c}{Obs. 1, class $\nu$}\\
\hline
 I$_\nu$  &  0.7 (frozen)    &   & 33$_{-4}^{+6}$ & 0.6$\pm0.1$ & 1.49 (88) &  & 1.4 \\
 II$_\nu$     & 1.33$\pm0.09$ & 12.1$_{-1.7}^{+2.2}$ & 14$_{-3}^{+8}$ & 1.5$_{-0.7}^{+0.6}$ & 1.42 (44) & 2.42 & 1.9\\
 III$_\nu$       & 1.5$_{-0.2}^{+0.3}$ & 7.6$_{-0.6}^{+2.0}$ & 90$_{-12}^{+57}$ & 0.01 (frozen)& 1.30 (31) & 2.66&  0.86 \\
 IV$_\nu$       & 2.10$_{-0.07}^{+0.06}$ & 6.2$\pm0.3$ & 110$_{-14}^{+17}$ & 0.01 (frozen)& 1.52 (68) & 7.9&  1.5 \\
\hline
\multicolumn{8}{c}{Obs. 5, class $\beta$}\\
\hline
 I$_\beta$       & 0.8$\pm0.2$ & 28$_{-14}^{+64}$ & 15$_{-5}^{+32}$ & 1.5$_{-1.2}^{+0.5}$ & 1.11 (43) & 0.77 & 1.2\\
 II$_\beta$       & 0.9$\pm0.4$ & 21$_{-13}^{+100}$ & 181$_{-31}^{+65}$ & 0.01 frozen & 1.20 (39) & 0.89 & 1.6\\
 III$_\beta$      & 1.5$\pm0.1$ & 6.2$_{-1.4}^{+1.2}$ & $>124$\tablenotemark{b} & 0.01 frozen & 1.13 (31) &  1.4 &  0.24 \\
\hline
\multicolumn{8}{c}{Obs. 2, class $\lambda$}\\     
\hline
Int. & kT$_{disk}$ & R$_{in}$\tablenotemark{a} & \multicolumn{2}{c}{Power law index} & $\chi_\nu^2$ & \multicolumn{2}{c}{unabs. fluxes}\\
     & (keV)       &  ($\times\frac{\mathrm{R}_\mathrm{G}}{\sqrt{cos(i)}}$ )& \multicolumn{2}{c}{$\Gamma$} & (d.o.f.) & 3-50~keV disk & 3-50~keV PL\\ 
     &             &                           &        &        &              & \multicolumn{2}{c}{($\times10^{-8}$~\ergcms)}\\
\hline
 I$_\lambda$      & \nodata & \nodata &  \multicolumn{2}{c}{2.51$_{-0.11}^{+0.09}$} & 1.14 (48) & & 0.95 \\
 II$_\lambda$     & \nodata & \nodata &  \multicolumn{2}{c}{2.6$\pm0.2$} & 0.95 (37) & & 1.2  \\
 III$_\lambda$       & 1.83$\pm0.09$ & 7$_{-2}^{+1}$ & \multicolumn{2}{c}{2 frozen} & 0.88 (29) & 4.4 & $<0.29$\\
 IV$_\lambda$       & 2.11$_{-0.07}^{+0.08}$ & 6$\pm1$ & \multicolumn{2}{c}{2 frozen} & 0.88 (31) &7.3 &  $<0.43$\\
 V$_\lambda$       & 2.05$\pm0.07$ & 5.1$_{-0.5}^{+0.9}$ & \multicolumn{2}{c}{2 frozen} & 1.08 (33) & 4.8 & $<0.49$  \\
 VI$_\lambda$       & 1.97$\pm0.06$ & 5.9$\pm0.6$ & \multicolumn{2}{c}{2.9$\pm0.3$} & 1.49 (41) & 5.4 & 1.9 \\
\hline
\hline 
\end{tabular}
\tablenotetext{a}{We take a correction factor f=1.9 \citep{shimura95}, M=14$M_\odot$, and D=10 kpc to convert the model normalization to 
R$_\mathrm{G}$ , R$_{in}=0.859 \times \sqrt{K} \times\frac{\mathrm{R}_\mathrm{G}}{\sqrt{cos(i)}}$, with K the model normalization and i the 
inclination angle.}
\tablenotetext{b}{The limit on this parameter is given at the 90\% confidence level.}
\label{tab:fits}
\end{table}

\newpage
\begin{table}
\caption{Best parameters obtained from the fits to the JEM-X+ISGRI (and PCA+HEXTE when 
available) spectra of Obs.~4/class $\chi$. The best fit model is 
{\tt{phabs(comptt+ga+powerlaw)}} in {\tt{XSPEC}} notation. The errors on the parameters are given at 
the 90$\%$ confidence level.}
\begin{tabular}{cccccccc}
\hline
\hline
Interval & kT$_{seed}$ & kT$_e$ & $\tau$ &$\Gamma$  & $\chi^2_{\nu}$& \multicolumn{2}{c}{2--200~keV unabs. fluxes\tablenotemark{a}}\\
  &  (keV) & (keV) &     &    & (d.o.f.) & \multicolumn{2}{c}{($\times10^{-8}$~\ergcms)} \\ 
   & &  &          &    &       & comptt & PL \\
\hline
 I$_\chi$\tablenotemark{b} & 0.32$\pm0.01$ & 6.66$\pm0.03$ & 2.084$\pm0.008$ & 2.62$\pm0.01$ & 1.60 (122) & 2.70 & 0.77\\
 II$_\chi$     & 0.33$_{-0.03}^{+0.06}$ & 6.66$_{-0.07}^{+0.34}$ & 2.22$_{-0.02}^{+0.2}$ & 2.74$_{-0.04}^{+0.02}$ & 0.92 (52) & 2.65  & 1.09\\
 III$_\chi$     & 0.26$_{-0.2}^{+0.04}$ & 7.0$\pm0.1$ & 2.16$_{-0.02}^{+0.03}$& 2.81$_{-0.07}^{+0.02}$ & 1.25 (51) & 2.32  & 1.24\\
\hline
\hline
\end{tabular}
\tablenotetext{a}{The fluxes are those from \integral/ISGRI (extrapolated to 2--200~keV). }
\tablenotetext{b}{Fits of this interval also include simultaneous \rxte\ data. Note that if the normalization constant
is frozen to 1.0 for PCA, the one we obtain from the fit for ISGRI is 0.952 and that for JEM-X 1.016.}
\label{tab:fitschi}
\end{table}
 
\newpage

\end{document}